**Multimodal azimuthal oscillations in electron beam generated E × B plasma**

Nirbhav Singh Chopra[1,2], Mina Papahn Zadeh[3], Mikhail Tyushev[3], Andrei Smolyakov[3], Alexandre Likhanskii[4], and Yevgeny Raitses[1]

[1] Princeton Plasma Physics Laboratory, Princeton University, Princeton, NJ 08543, United States of America

[2] Department of Astrophysical Sciences, Princeton University, Princeton, NJ 08544, United States of America

[3] Department of Physics and Engineering Physics, University of Saskatchewan, Saskatoon SK S7N 5E2, Canada

[4] Applied Materials Inc, 35 Dory Rd, Gloucester, MA 01930, United States of America

**Abstract**

Electron beam (e-beam) generated plasmas with applied cross electric and magnetic (E × B) fields are promising for low-damage material processing. However, these plasmas can be subject to the formation of azimuthally propagating structures that enhance the radial transport of energetic charged species, which can harm the gentle processing capability of the plasma. In this work we investigate the azimuthal structure formation in an e-beam generated E × B plasma using experimental diagnostics and 2D3V particle-in-cell simulations. Our findings demonstrate the formation of multiple simultaneously occurring azimuthally propagating modes that exhibit a nontrivial radial dependence. It is suggested that the multimodal azimuthal spectrum is caused by the complex nature of the ion dynamics in the plasma.

## I. Introduction

Electron beam (e-beam) generated plasmas with applied electric and magnetic (E × B) fields are promising for applications requiring efficient generation of ions and radicals in low pressure environments [1–4]. A key capability of magnetically confined e-beam plasma sources is that they can generate reactive species while maintaining low energetic particle flux to substrates placed in the periphery of the plasma region [5–7].

In many axisymmetric E×B plasma systems, such as Penning discharges, Hall discharges, sputtering magnetrons, and Bernas sources, electrons are magnetized (electron Hall parameter $\Omega_e \gg 1$), while ions are usually non-magnetized (ion Hall parameter $\Omega_i \ll 1$) or very weakly magnetized ($\Omega_i \lesssim 1$) [8–12]. Therefore, ion transport is dominated by an ambipolar electric field and collisions with neutrals. Previous work on e-beam generated E×B Penning plasma has demonstrated the formation of an ion-confining radial electric potential well [1,8,9]. This potential can be leveraged for low-damage threshold material processing applications by limiting energetic ion flux, thereby preventing energetic ions from damaging material surfaces.

However, several operating regimes of axisymmetric $E \times B$ plasma devices are accompanied by large-scale azimuthally propagating fluctuations in plasma density and potential ('spokes') induced by the modified Simon-Hoh instability (MSHI) [8,9,12–14]. The spoke has been shown to energize ions in the radial direction, thereby reintroducing energetic ions to the peripheral substrate processing region [15]. Therefore, mitigating the formation of the spoke may enable a wider operating regime of gentle processing capability of the e-beam generated $E \times B$ plasma.

In this work, we experimentally and computationally investigate the structure and formation of azimuthally propagating plasma structures in an e-beam generated $E \times B$ plasma. Our experimental findings indicate the existence of two regions with distinct azimuthal mode formation. We also performed 2D3V Particle-in-cell (PIC) simulations in the plane perpendicular to the applied magnetic field to independently determine the radial macroscopic profiles and azimuthal mode spectrum. Excellent agreement is shown between experimental findings and the simulation.

The paper is organized as follows: Section II discusses the experimental setup of the e-beam generated $E \times B$ plasma. Section III discussed the plasma diagnostics and measurement procedures used, as well as the PIC simulation setup. Section IV details the results of the probe measurements and their analysis, as well as the experimental validation of the PIC simulations. Analytical models of the radial electron and ion transport and azimuthal instability onset are proposed and compared to the experimental measurements in Section V. Conclusions are summarized in Section VI.

## II. Experimental setup

The experimental setup consisted of an e-beam generated $E \times B$ plasma chamber with a conducting anticathode installed on the axially opposite side of the chamber from the cathode. The setup here is nearly identical to the setup described in Ref. [16]. However, in the present setup the thermionic cathode was significantly modified, which is depicted in Figure 1. The tungsten filament with exposed length 10 mm and diameter 0.4 mm is inserted through a stainless-steel cathode plate with radius $R_{c,\text{plate}} = 4.5$ cm. The filament is electrically isolated from the cathode plate with ceramic tubes, so that heater current does not pass directly through the cathode plate body. Both the thermionic filament and the cathode plate are biased to the applied cathode bias voltage $V_c$. The modified cathode increases the region of axial electron confinement to the radial extent of the cathode plate $R_{c,\text{plate}}$ [16].

In this work, the anticathode is biased to the cathode potential such that is electron repelling, i.e. operated in 'repeller mode' with anticathode bias potential $V_{atc} = V_c$. This is in contrast to 'collector mode', whereby the anticathode is biased to the anode potential $V_{atc} = 0$ V and therefore collects nearly all incident electrons. The effect of the anticathode operating in collector mode is not considered in the present work.

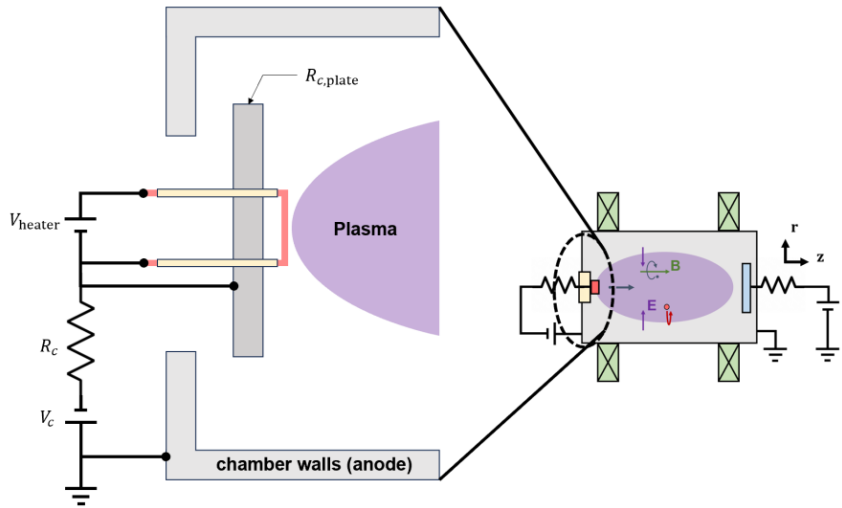

**Figure 1.** Schematic of modified thermionic cathode installed on e-beam generated $E \times B$ chamber.

## III. Methods
### 1. Diagnostics and measurement procedure

A Langmuir probe (LP) and emissive probe (EP) diagnostic were used to determine the electron energy distribution function (EEDF), electron density, electron temperature, and plasma potential at different radii. These diagnostics are described in further detail in Ref. [16].

The coherent azimuthal plasma density oscillations were determined by an azimuthally oriented ion probe array (Figure 3). The ion probe array consists of two negatively biased tungsten wires with identical exposed collecting areas, each with diameter 1.1 mm and exposed wire length of 1.7 mm. Each tungsten wire is inserted into a ceramic tube of outer diameter 2.1 mm. The probe tips are separated by a distance $\Delta x = 2.1$ cm such that the azimuthal angle subtended by the probe array is given by $\theta_{12} = 2\arctan(\Delta x/2r_0)$, where $r_0$ is the probe array insertion depth. As a result, the radial distance of the probe tips to the geometric center of the plasma is given by $r = \sqrt{\left(\frac{\Delta x}{2}\right)^2 + r_0^2}$. The geometric properties of the probe array shown in Table I.

**Table I.** Geometric properties of the probe array.

| $r_0$ [cm] | $r$ [cm] | $\theta_{12}$ [rad] ([deg]) |
|---|---|---|
| 1.0 | 1.5 | 1.6 (93°) |
| 2.0 | 2.3 | 1.0 (55°) |
| 3.0 | 3.2 | 0.7 (39°) |

The tungsten wire is individually biased to a potential of $V_b = -100$ V relative to the grounded chamber walls, such that the probe ion collection is in an orbital motion limited (OML) regime. Under such conditions, the ion current $I_j$ collected by the $j$th ion probe ($j = 1$ or 2 corresponding to ion probe 1 (IP1), and ion probe 2 (IP2)) is independent of the electron temperature and is proportional to the local plasma density [17,18],

$$I_j(t) = \frac{2}{\sqrt{\pi}} e n_j(t) A_{pr} \sqrt{-\frac{e(V_b - V_{pl})}{2\pi m_i}}, \tag{1}$$

where $n_j(t)$ is the temporally fluctuating plasma density at the $j$th probe, $A_{pr}$ is the probe collection area, $V_{pl}$ is the plasma potential, and $m_i$ is the Argon ion mass. Here the plasma is assumed to be quasineutral, $n_j \approx n_i \approx n_e$ where $n_i$ and $n_e$ are the ion and electron densities respectively.

A two point-correlation technique is used to determine the azimuthally coherent portion of the fluctuating signal collected by each ion probe. The cross power spectral density (CSD) is used to quantitatively determine the coherence and relative phase of $n_1(t)$ and $n_2(t)$ [19,20]. Representative ion current signals collected by IP1 and IP2 and their cross spectral density $P_{12}$ are shown in Figure 2. The CSD is defined as the Fourier transform of the cross-correlation of the two signals [20],

$$\begin{aligned}P_{12}(f) &\equiv \int_{-\infty}^{\infty} \left[ \int_{-\infty}^{\infty} n_1(t)\, n_2(t+\tau) dt \right] e^{-i2\pi f\tau} d\tau \\ &\equiv F\left( \int_{-\infty}^{\infty} n_1(t)\, n_2(t+\tau) dt \right).\end{aligned} \tag{2}$$

where $f$ is the frequency. The cross-coherence is defined as [20]

$$C_{12}(f) \equiv \frac{|P_{12}|^2}{|P_{11}||P_{22}|}, \tag{3}$$

which quantifies the fraction of spectral power transferred from $n_1$ to $n_2$, for each spectral component $f$. Moreover, the argument of the CSD, $\text{Arg}(P_{12})$, quantifies the relative phase between $n_1$ and $n_2$ for each spectral component $f$. Here the argument function is taken to have the convention that for a complex valued function $g$, $\pi < \text{Arg}(g) \leq \pi$.

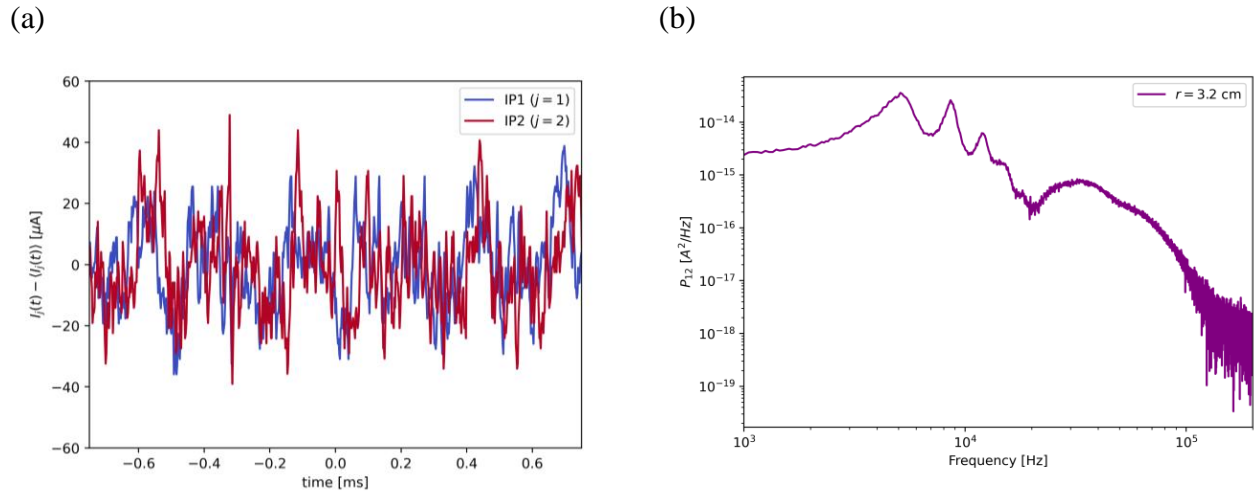

**Figure 2.** (a) Ion current temporal signal $I_j(t)$ determined by IP1 ($j = 1$) and IP2 ($j = 2$) for $B = 100$ G, $V_c = 55$ V, $p = 0.1$ mTorr, with mean signal subtracted (b) Cross spectral density $P_{12}$ of $I_1$ and $I_2$.

Azimuthal oscillations were also independently determined by a fast frame camera. A Phantom v7.3 high speed camera operating at 95 kfps imaged the optical emission of the plasma in the $r - z$ plane (Figure 3). The fast frame images were then averaged along the axial (z) direction and Fourier transformed in time to determine the radial dependence of the plasma emission oscillations.

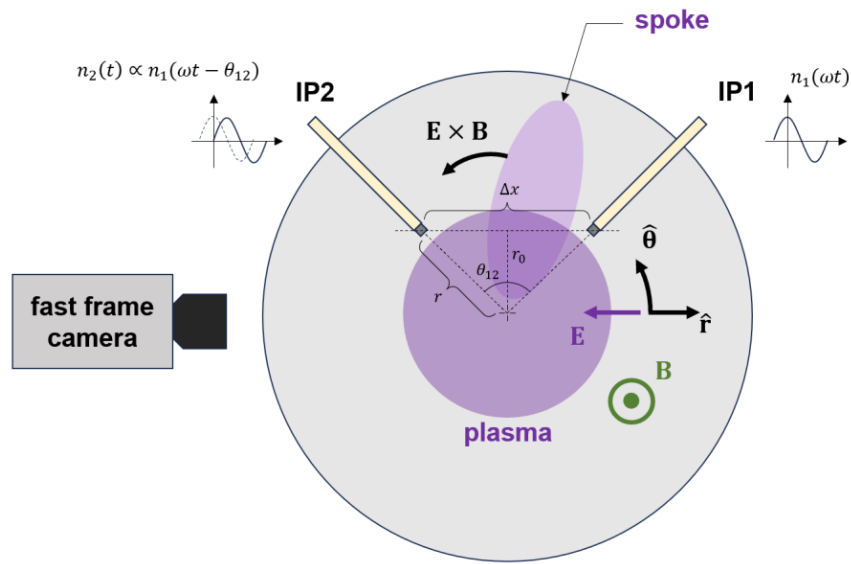

**Figure 3.** Schematic of diagnostic setup for azimuthal oscillation characterization.

## 2. Particle in-cell method

A 2D3V particle-in-cell method (PIC) was developed using the WarpX code [21] to simulate the electron beam generated E × B plasma in the plane perpendicular to the applied magnetic field. The simulation domain is indicated in Figure 4, and relevant simulation parameters are shown in Table II. Simulations were performed to most closely emulate the conditions of the experimental setup. The major difference between the PIC setup and the experimental setup is the radius of the anode in the computational domain, $R_{sim.} = 5$ cm as compared to the experimental chamber radius of 10 cm. Azimuthal oscillations and time averaged radial profiles of electron density, temperature, and plasma potential are determined from the PIC simulations and are compared to the experimental measurements.

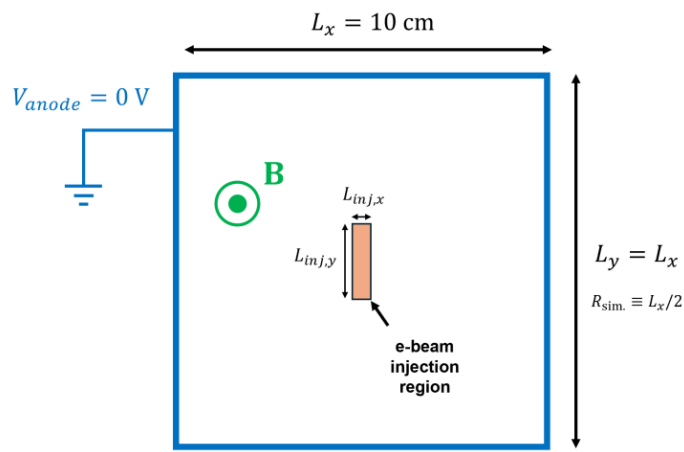

**Figure 4.** Diagram of simulation domain.

**Table II.** Properties of the PIC method used in this work.

| Parameter | Value |
|---|---|
| $\Delta x$ | 0.130 mm |
| $\Delta t$ | $1.7 \times 10^{-11}$ s |
| C | 0.26 |
| $N_{macro.}$ | $1.0625 \times 10^7$ |
| $R_{sim.}$ | 5 cm |
| $L_{inj,x}$ | 0.4 mm |
| $L_{inj,y}$ | 10 mm |
| $I_{inj}$ | 100 mA |
| $p$ | 0.1 mTorr (Argon) |
| $B$ | 100 G |

## IV. Results
### 1. Experimental measurements

The Fourier spectra of the plasma density fluctuation $F(n_1(t))$ at $r = 1.5$ cm and 3.2 cm are shown in Figure 5, for operating conditions $V_c = -55$ V, $B = 100$ G, $p = 0.1$ mTorr. Both $r = 1.5$ cm and $r = 3.2$ cm have prominent low frequency modes occurring at $f_0 = 5 - 7$ kHz and the first harmonic $f_1 \approx 2f_0$, with $f_1$ increasing in amplitude at $r = 3.2$ cm. There is also a higher frequency mode $f_H = 30$ kHz that is more prominent at $r = 1.5$ cm.

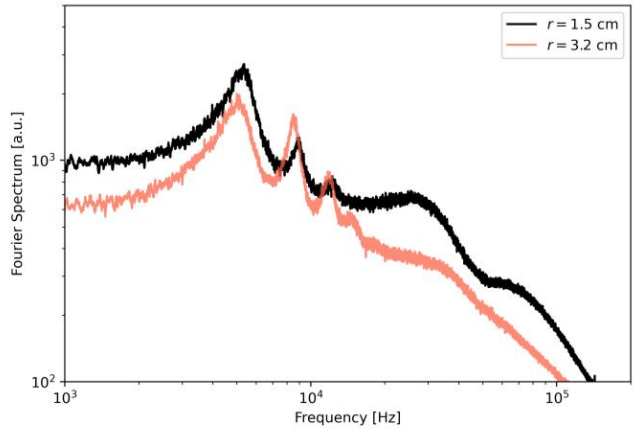

**Figure 5.** FFT of ion density temporal trace determined by IP1, measured at $r = 1.5$ cm and $r = 3.2$ cm. Parameters are $B = 100$ G, $V_c = 55$ V, $p = 0.1$ mTorr.

The cross coherence of $n_1(t)$ and $n_2(t)$, $C_{12}(f)$ is shown in Figure 6, for the same experimental parameters. Figure 6 a and b indicate that at $r = 1.5$ cm both low and high frequency modes at $f_0$ and $f_H$ have a high coherence with $C_{12} > 0.5$. Moreover, both modes obey $\mathrm{Arg}(P_{12}(f_0)) \approx \mathrm{Arg}(P_{12}(f_H)) \approx -\theta_{12}$, indicating both $f_0$ and $f_H$ are single lobed ($m = 1$) modes that are azimuthally propagating in the $+\mathrm{E} \times \mathrm{B}$ direction. At $r = 3.2$ cm, the coherence of the low frequency mode remains high $C_{12}(f_0) > 0.8$, while the high frequency mode coherence reduces $C_{12}(f_H) < 0.5$. The azimuthally propagating modes occurring at $f_0$ and $f_H$ are also corroborated by the fast frame imaging diagnostic (Appendix 2).

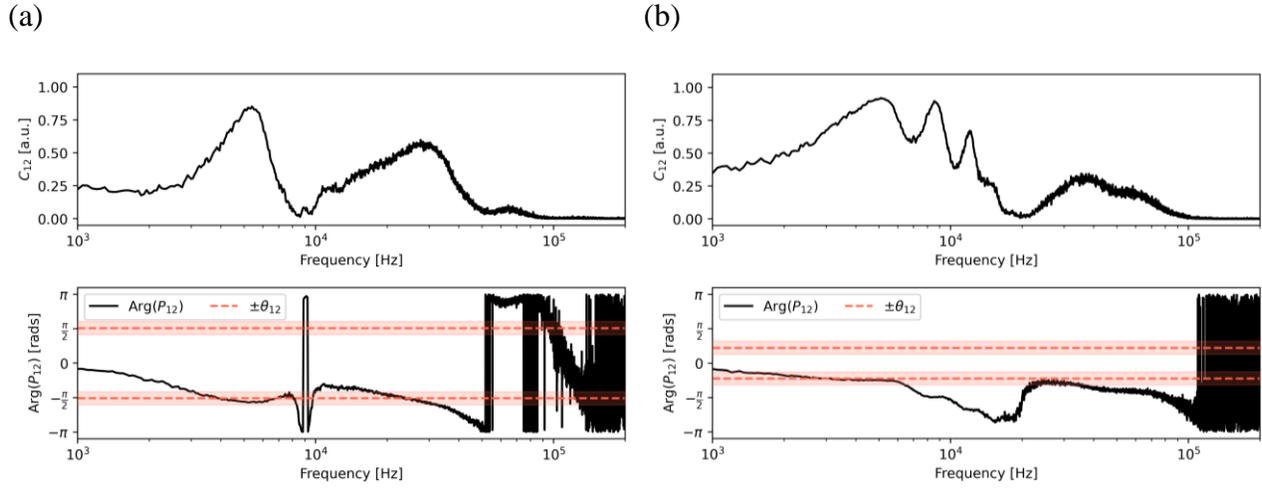

**Figure 6.** Cross coherence and argument of cross spectral density of ion density temporal traces, measured at (a) $r = 1.5$ cm, (b) $r = 3.2$ cm. Parameters $B = 100$ G, $p = 0.1$ mTorr, $V_c = -55$ V.

The cross-coherence dependence on the applied magnetic field is shown for $r = 1.5$ cm and $r = 3.2$ cm in Figure 7. A trend of enhancement in the coherence of $f_0$ and $f_1$ and suppression of $f_H$ can be observed for increasing magnetic field. This trend is also confirmed by the fast frame image analysis (Appendix 2).

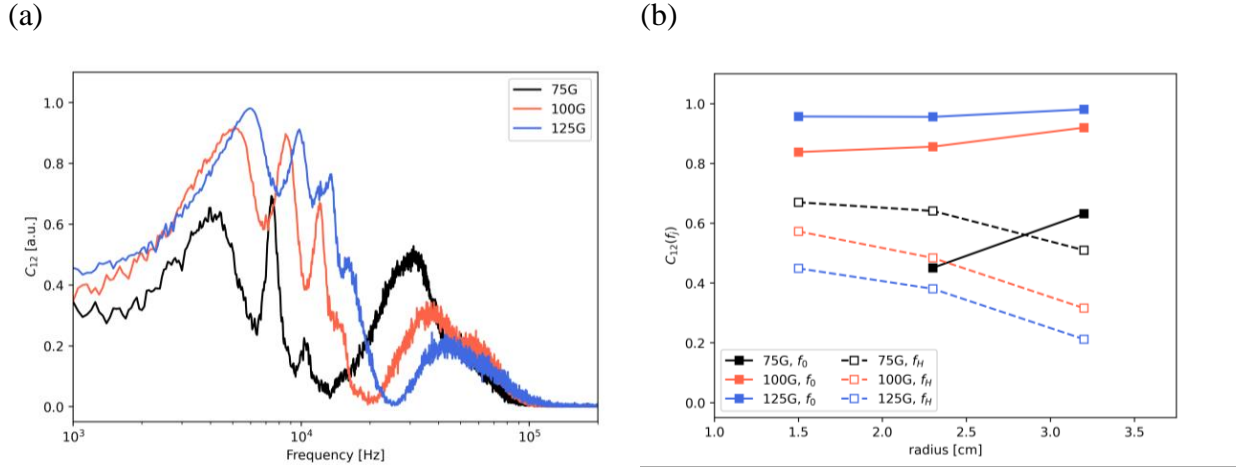

**Figure 7.** (a) Ion probe cross coherence spectra measured at $r = 3.2$ cm for several values of the applied magnetic field (b) Mode cross coherence vs. radial coordinate for the low frequency mode $f_0$ and high frequency mode $f_H$ for several magnetic fields. Experimental parameters are $p = 0.1$ mTorr, $V_c = -55$ V.

## 2. Experimental validation of 2D PIC model

A snapshot of the ion density computed by the PIC simulation after reaching quasi-steady state is shown in Figure 8. Also shown are the azimuthal and radial components of the wavevector associated with the observed density wave, $\mathbf{k} = k_r \hat{r} + k_\theta \hat{\theta}$. Indeed an azimuthally rotating plasma density structure is observed propagating in the $+\text{E} \times \text{B}$ direction. Furthermore, the simulation indicates that the rotating spoke is accompanied by both radially outward and inward propagating wavevectors components: ahead of the spoke, $k_r < 0$, while behind the spoke, $k_r > 0$.

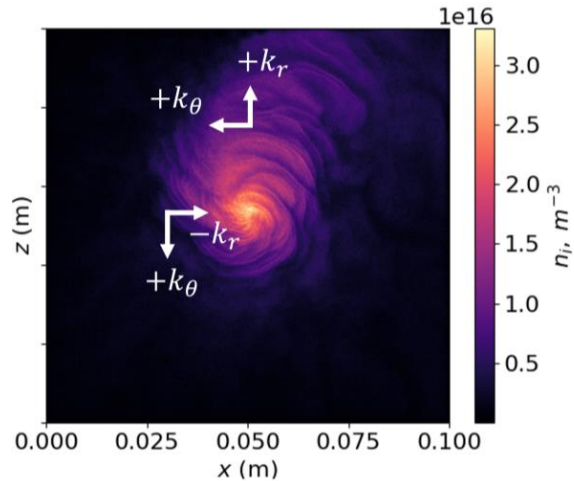

**Figure 8.** Snapshot of ion density after simulation reaches saturation. Also indicated are characteristic azimuthal and radial components of the wavevector.

A comparison of radial profiles of electron density, electron temperature, plasma potential, and azimuthal pseudospectra determined experimentally and from the PIC simulations are shown in Figure 9. The normalized radial coordinate is defined as $\rho = r/R_{\text{sim.,expt.}}$, where $R_{\text{sim.}} = 5$ cm and $R_{\text{expt.}} = 10$ cm are the anode radius in the simulation and experiment, respectively. The time averaged electron density, electron temperature, and plasma potential radial profiles all show good agreement.

The azimuthal modes occurring in the PIC simulation are determined using the MUSIC method [15]. The PIC simulation reproduces the occurrence of multiple coherent azimuthal modes. However, the frequency of the azimuthal modes in the simulation are larger than in experiment by approximately a factor of 2. This may be due to the radius of the computational domain being smaller than the experimental chamber radius by a factor of 2. Indeed, previous works have demonstrated increasing computational domain size reduces the frequency of azimuthal oscillations in e-beam generated $\text{E} \times \text{B}$ plasmas [15].

Quantitative agreement in the mode peak frequencies is observed between the simulated MUSIC spectrum and the experimentally determined cross-coherence when the frequency transformation $f' = f \cdot (R_{\text{expt.}}/R_{\text{sim.}})$ is applied to the experimental cross-coherence. This transformation is physically meaningful, since the smallest nonzero azimuthal wavevector that can occur in a system of radius $R_0$ is $k_\theta^0 = 2/R_0$. Therefore, a frequency upscaling by a factor of $k_\theta^{0,\text{sim.}}/k_\theta^{0,\text{expt.}} = R_{\text{expt.}}/R_{\text{sim.}} = 2$ may occur in the simulation MUSIC spectrum compared to the experimentally observed azimuthal spectrum. This speculation should be further investigated by performing PIC simulations with a computational domain radius equal to the experimental chamber radius.

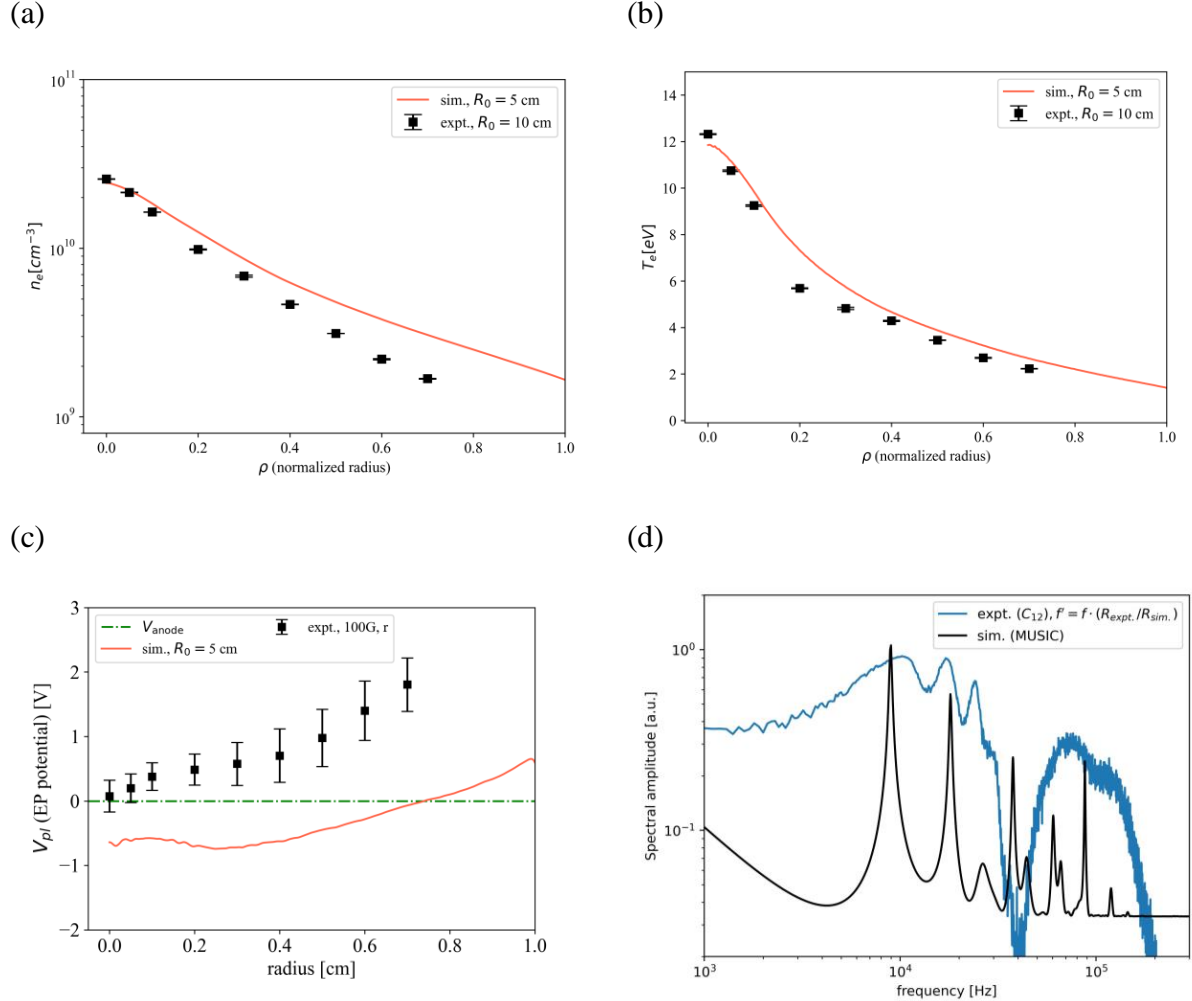

**Figure 9.** Validation of radial profiles determined by experiment (expt.) and simulation (sim.) for (a) Electron density $n_e$, (b) Electron temperature $T_e$, and (c) Plasma potential $V_{pl}$. Parameters are $B = 100$ G, $p = 0.1$ mTorr, $V_c = -55$V. (d) Density fluctuation azimuthal pseudospectra for $\rho = 0.32$ (expt.: cross-coherence $C_{12}$, sim.: MUSIC spectrum). The experimental cross-coherence spectrum frequency has been scaled by the ratio $R_{\text{expt.}}/R_{\text{sim.}}$.

This agreement of simulation with experimentally determined radial profiles and azimuthal mode spectrum indicates that the presently considered regime of the e-beam generated E × B plasma operated with an electron repelling anticathode is an inherently 2D3V system.

## V. Discussions
### 1. Radial electron and ion transport

Here we apply a 1-dimensional model to determine the cross field anomalous electron transport in the radial direction, which is based on the electron continuity equation. This model is closely related to the 0-dimensional model developed in Ref. [16]. Additionally, we apply a 1-dimensional model of the ion continuity equation to determine the radial outflow velocity of ions, which is used further in Section V.2 to determine the onset of the modified Simon-Hoh instability. The electron and ion continuity equations are given by

$$\nabla \cdot \mathbf{\Gamma}_e = R_{iz}, \tag{4}$$

$$\nabla \cdot \mathbf{\Gamma}_i = R_{iz}, \tag{5}$$

where $\mathbf{\Gamma}_e$ and $\mathbf{\Gamma}_i$ are the local electron and ion fluxes respectively and $R_{iz}$ is the volumetric ionization rate given by

$$R_{iz} = n_g \left(\frac{2}{m_e}\right)^{1/2} \int_0^\infty d\varepsilon \, \sigma_{iz}(\varepsilon) \varepsilon^{\frac{1}{2}} f_e(\varepsilon), \tag{6}$$

with the gas density $n_g = p/kT_g$ and neutral gas temperature assumed to be $T_g = 300\text{K}$. Integrating Eqs. (4) and (5) over a cylindrical control volume of length $L_{ch}$ and radius $r$, and solving for the radial flux, we find

$$\bar{\Gamma}_e^r(\alpha) = \bar{\Gamma}_e^b + \bar{R}_{iz} - \bar{\Gamma}_{e,\text{loss}}^z, \tag{7}$$

$$\bar{\Gamma}_i^r = \bar{R}_{iz} - \bar{\Gamma}_i^z, \tag{8}$$

where $\bar{\Gamma}_{e,\text{loss}}^z$, $\bar{\Gamma}_e^r(\alpha)$, $\bar{\Gamma}_e^b$, $\bar{R}_{iz}$, $\bar{\Gamma}_i^z$ and $\bar{\Gamma}_i^r$ are the volume integrated axial electron loss flux, radial electron flux, injected electron beam flux, ionization rate, axial ion flux, and radial ion flux respectively. The semi-empirically determined anomalous parameter $\alpha$ is the inverse Hall parameter. The quantities expressed in Eqs. (7) and (8) are explicitly defined as

$$\bar{\Gamma}_{e,\text{loss}}^z = 2\pi \int_0^r r' dr' \frac{n_e v_{the}}{4} \left\{ \exp\left(-\frac{(V_{pl} - V_c)}{kT_e}\right) + \min\left[\exp\left(-\frac{(V_{pl} - V_{atc})}{kT_e}\right), 1\right] \right\}, \tag{9}$$

$$\bar{\Gamma}_e^r(\alpha) = -\mu_{e\perp}(\alpha)\left(n_e E_r + \frac{\nabla p_e}{e}\right) \cdot (2\pi L_{ch} r), \tag{10}$$

$$\bar{R}_{iz}(r) = 2\pi L_{ch} \int_0^r r' dr' R_{iz}, \tag{11}$$

$$\bar{\Gamma}_i^z = 2 \times 2\pi \int_0^r r' dr' (0.61 n_e v_{\text{Bohm}}), \tag{12}$$

and the injected electron beam flux is approximately the discharge current $\bar{\Gamma}_e^b \approx I_d/e$. Here we make the approximations that the net axial flux of electrons is given by the electron thermal flux confined by the cathode and anticathode sheaths. We also assume axial uniformity, that the radial electron flux is given by the drift-diffusion approximation, and that the EEDF is isotropic. Furthermore, we assume the cathode and anticathode sheaths are planar and therefore approximate the axial ion flux as the Bohm current, with the sheath density $n_s = 0.61\, n_e$ and ion velocity entering the sheath as $v_{\text{Bohm}} = (k_B T_e/m_{Ar})^{0.5}$ where $m_{Ar}$ is the mass of an Argon ion. The prefactor of 2 occurring in Eq. (12) accounts for Bohm ion current collected by both the anticathode and cathode surfaces. Finally, we neglect the Simon effect for electrons 'short circuiting' across the magnetic field lines through the axial conducting boundaries [22,23]. We make this assumption since the discharge is operated in repeller mode, in which case both anticathode and cathode have a large electron reflecting sheath. Under such conditions negligible electron flux will strike either conducting boundary.

In a previous work on a similar e-beam generated $E \times B$ plasma, it was determined that electron-neutral collisions are insufficient to explain the electron cross field transport, and electron transport rather is so-called anomalous, i.e. driven by effective electron collisions with fluctuations in the plasma [16]. In the case of anomalous cross field electron transport such that $\nu_{en} \ll \alpha \omega_{ce}$, the electron cross-field mobility can be approximated as $\mu_{e\perp}(\alpha) \approx \alpha/B$ [24]. Under such conditions, Eq. (7) can be solved for the anomalous parameter,

$$\alpha(r) = \frac{\bar{\Gamma}_e^b + \bar{R}_{iz} - \bar{\Gamma}_{e,\text{loss}}^z}{-\frac{1}{B}\left(n_e E_r + \frac{\nabla p_e}{e}\right) 2\pi r L_{ch}}. \tag{13}$$

The anomalous parameter $\alpha$ as a function of the radial coordinate is shown for several values of the applied magnetic field in Figure 10. The statistically determined anomalous parameter using the mean-squared ion density fluctuation determined from IP1, $\alpha_{YR} = \frac{\pi}{4}\frac{\langle(n-n_0)^2\rangle}{n_0^2}$, as derived by Ref. [24], is also shown.

Both the model based on Eq. (13) and the statistically determined $\alpha_{YR}$ agree in the trend of relatively low value of the anomalous transport factor in the region $r \leq 2$ cm, with sharply increasing anomality for $r > 2$ cm. Additionally, both $\alpha_{\text{mod.}}$ and $\alpha_{YR}$ tend to increase with increasing magnetic field, indicating higher cross field electron transport at larger magnetic fields. This trend predicted by the model agrees with the observation of enhanced mode coherence at larger magnetic field observed from the azimuthal ion probe measurements (Figure 7). Therefore, the enhancement of anomalous transport in the $r > 2$ cm is likely due to the enhancement of the spoke.

The modeled anomalous parameter is reduced significantly by at least a factor of 2 at $r = 1.5$ cm compared to $r = 3.2$ cm. This is caused by the enhanced ionization rate near the discharge center for $r \leq 2$ cm. For $r > 2$ cm, ionization is not sufficient to sustain the electron transport, and therefore the anomalous factor increases to maintain the electron continuity.

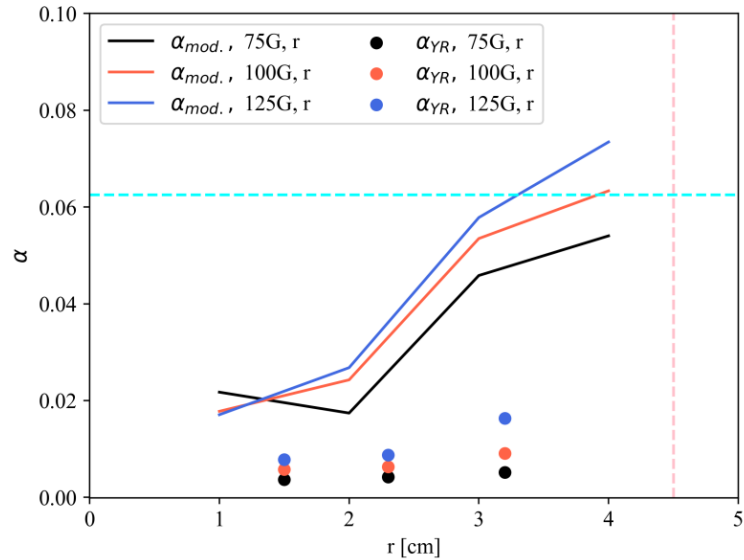

**Figure 10.** Anomalous parameter $\alpha_{\text{mod.}}$ as a function of radius, as determined by Eq. (13), for $B = 75, 100,$ and $125$ G. Also shown is the statistically determined anomalous parameter $\alpha_{YR}$. The cathode and anticathode radius $R_{c,\text{plate}} = R_{atc} = 4.5$ cm is indicated by the vertical dashed line, and the value of the anomalous parameter determined by Bohm is indicated by the horizontal dashed line ($\alpha = 1/16$, Ref. [25]).

To explain the radial dependence of the anomalous transport, the radial dependence of the electron current terms from Eq. (7) are shown in Figure 11 in the case $B = 100$ G. For $r \geq 2$ cm the electron current due to ionization $e\bar{R}_{iz}$ saturates. However, the absolute value of the gradient terms in the denominator of Eq. (7) continue to reduce for $r \geq 2$ cm. The dominant term in the denominator is the electron pressure gradient $\nabla p_e$, which indeed shows a sharp decline in magnitude for $r > 2$ cm. Therefore, an enhancement in $\alpha$ for $r \geq 2$ cm occurs to compensate for the reduction in the electron pressure gradient, so that the radial electron current $e\bar{\Gamma}_e^r$ can balance the injected beam current $e\bar{\Gamma}_e^b$.

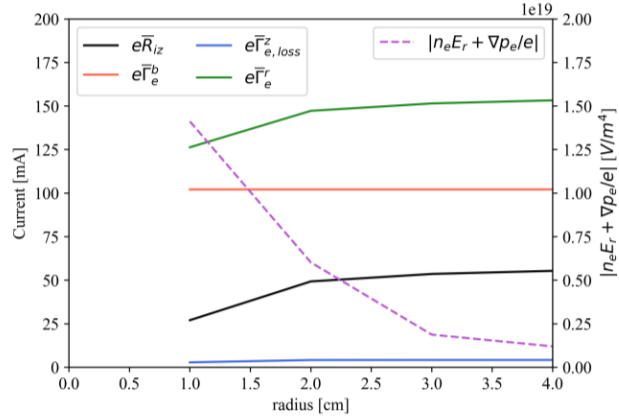

**Figure 11.** Radial dependence of electron current terms and denominator from Eq. (7), for the case $B = 100$ G.

Using the experimentally determined radial profiles for the electron density, electron temperature, and ionization rate (Appendix 1), the radial ion current can be determined by solving Eqs. (8), (11), and (12) for $e\bar{\Gamma}_i^r$,

$$e\bar{\Gamma}_i^r = 2\pi L_{ch} \int_0^r r'dr' R_{iz} - 2 \times 2\pi \int_0^r r'dr'(0.61 n_e v_{\text{Bohm}}). \tag{14}$$

Additionally, the radial outward ion velocity can be computed as

$$\bar{v}_{0i} = \bar{\Gamma}_i^r / 2\pi r L_{ch}. \tag{15}$$

The resulting values of $e\bar{\Gamma}_i^r$ and $\bar{v}_{0i}$ are shown in Figure 12. Evidently a radially outward ion current is necessary to satisfy global ion current continuity, generating an ion flow velocity that reaches approximately $\bar{v}_{0i} = 350$ m/s in the radial outward direction. For $r > 2$ cm, the radial ion current and resulting radial ion velocity reduce. This occurs since the volume integrated ionization rate saturates for $r \geq 2$ cm (Appendix 1, Figure 17a), while the surface integrated Bohm ion current leaving the cylindrical endcaps to either cathode and anticathode continues to increase linearly (Appendix 1, Figure 18). This steady state radial outflow of ions has been predicted in previous works of similar e-beam generated $E \times B$ plasmas [26]. Doppler shift laser induced fluorescence (LIF) measurements are currently being conducted to directly measure the ion flow in the plasma and will be reported in a separate paper.

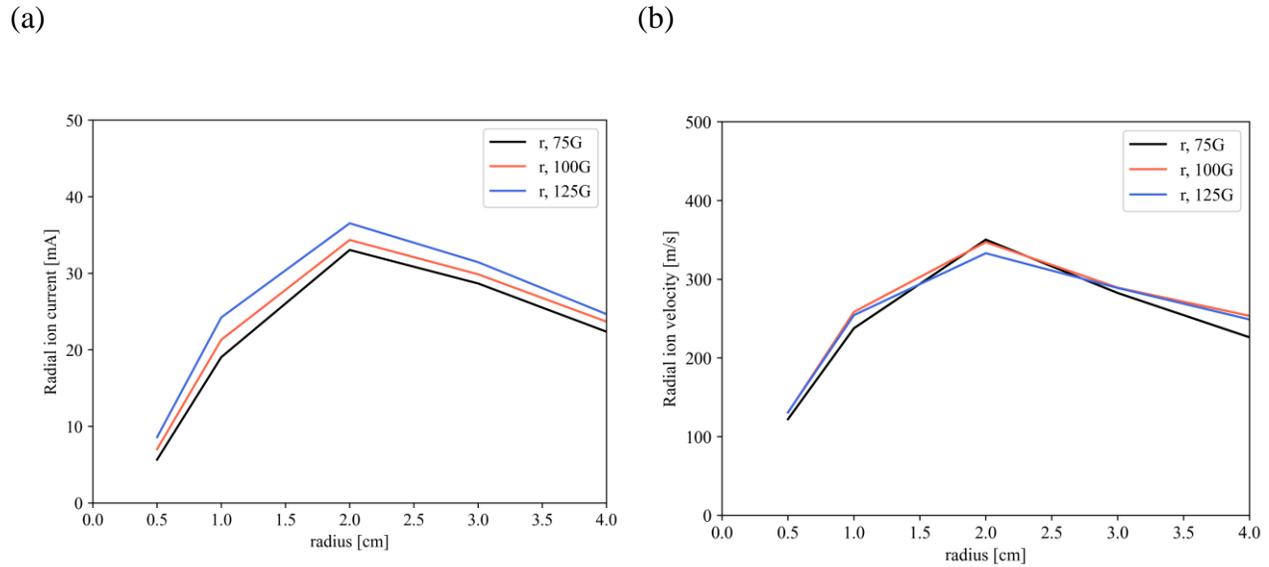

**Figure 12.** Radial dependence of (a) ion current $e\bar{\Gamma}_i^r$ and (b) radial ion velocity determined by ion current continuity, for magnetic fields $B = 75, 100,$ and $125$ G.

## 2. Physical origins of multimodal azimuthal oscillations

The magnetic field scaling of the experimentally observed azimuthal modes are shown in Figure 13. Moreover, the $f_0$ mode is amplified at larger radii $r \geq 2$ cm, while the $f_H$ mode decays at larger radii. The trend of increasing azimuthal modes frequency with increasing magnetic field is consistent with previous studies on similar E × B devices [8,9,15].

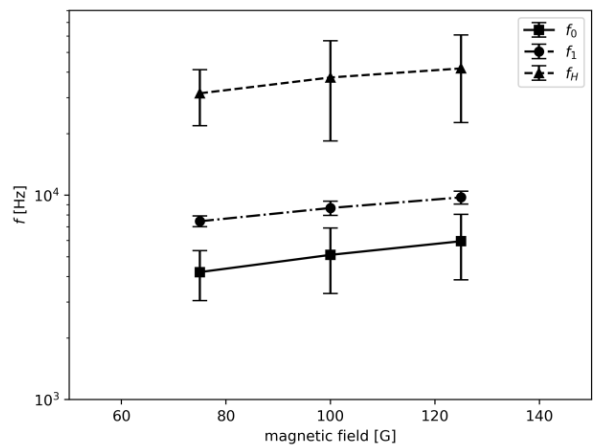

**Figure 13.** Frequency of $f_0, f_1,$ and $f_H$ modes at radial position $r = 3.2$ cm, as a function of applied magnetic field. Error bars in the mode frequency are given from the FWHM of the peaks in the cross-coherence spectrum. Experimental parameters are $B = 100$ G, $p = 0.1$ mTorr, $V_c = -55$V.

The modified Simon-Hoh dispersion relation derived from the linearized fluid equations is given as [27]

$$\omega = \omega_{0i} + \frac{k^2 c_s^2}{2\omega_*} + \sqrt{\frac{k^4 c_s^4}{4\omega_*^2} + \frac{k^2 c_s^2}{\omega_*}(\omega_{0i} - \omega_0)} \tag{16}$$

where $k = \left(k_r^2 + k_\theta^2\right)^{1/2}$ is the wavevector amplitude, $c_s = (k_B T_e/m_{Ar})^{0.5}$ is the bulk ion sound speed, $\omega_{0i} = \mathbf{k} \cdot \mathbf{v_{0i}}$ is the ion response frequency corresponding to ion flow with velocity $\mathbf{v_{0i}}$, $\omega_0 = -k_\theta E_r/B$ is the electron E × B frequency, and $\omega_* = -k_\theta k_B T_e/eBL_n$ is the electron diamagnetic drift frequency. Instability occurs for $\gamma \equiv \mathrm{Im}(\omega) > 0$. Importantly, electron E × B flow is always destabilizing, while the ion response is destabilizing when there is ion flow counterpropagating with the wavevector, i.e. $\mathbf{k} \cdot \mathbf{v_{0i}} < 0$.

There are several mechanisms in which ion flow can contribute to the MSHI growth rate [9]. Here we consider the effect of the radially outward directed equilibrium ion flow velocity estimated from the ion continuity Eq. (15), $\bar{v}_{0i} > 0$. This ion flow is destabilizing for the case $k_r < 0$. Indeed, the PIC simulations indicate the existence of a spatially dependent wavevector $\mathbf{k} = k_r \hat{\mathbf{r}} + k_\theta \hat{\boldsymbol{\theta}}$, with $k_r$ taking both positive and negative values (Figure 8). Therefore, these different processes associated with the ion flow may further destabilize the drift wave associated with Simon-Hoh Instability onset.

Using the experimental radial profiles for the electron density, electron temperature, and plasma potential, we determine radial profiles of $\gamma$ for the cases with and without ion responses. Here we assume a single lobed ($m = 1$) spoke centered at $R_{c,\mathrm{plate}}/2$, such that the azimuthal wavenumber is $k_\theta = 2/R_{c,\mathrm{plate}}$. Furthermore, in the cases where nonzero ion response is considered, the radial wavenumber is assumed to be singled lobed in the radial direction, such that $k_r = \pi/R_{c,\mathrm{plate}}$. However, as supported by the simulations (Figure 8), fine structures in the plasma density may form in the radial direction, indicating that the radial wavenumber may be significantly larger than considered here.

The calculated MSHI growth rate $\gamma$ radial profiles are shown in Figure 14, considering the cases of no ion response and bulk ion velocity outward $\bar{v}_{0i}$ as calculated by Eq. (15). Evidently, $\gamma$ shows good agreement with the $f_0$ and $f_1$ modes under assumption of no ion response and outward ion flow respectively.

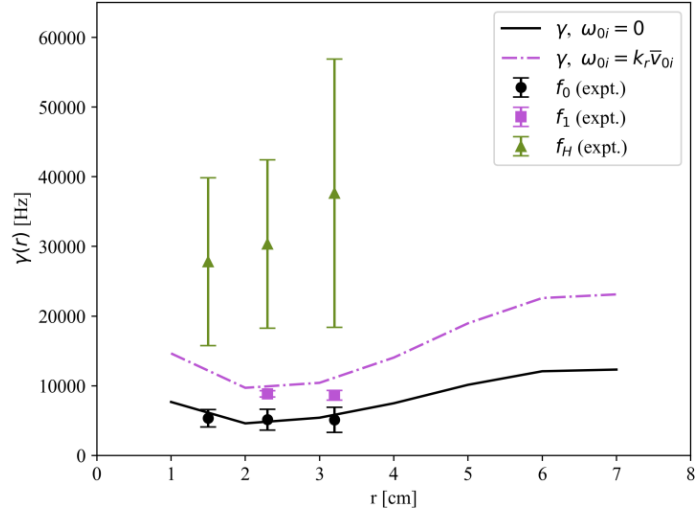

**Figure 14.** MSHI growth rate assuming $\omega_{0i} = 0$, $\omega_{0i} = k_r \bar{v}_{0i}$. Also shown are the experimentally determined $f_0$, $f_1$, and $f_H$ mode dependence on radius as determined by the azimuthal ion probe array. Experimental parameters are, $B = 100$ G, $p = 0.1$ mTorr, $V_c = -55$V.

This suggests future experimental studies and simulations should be carried out to precisely determine the ion dynamics and their role in the spoke formation in e-beam generated E × B plasmas, for instance, using Doppler shift LIF [28,29]. MSHI theory as described by Eq. (16) does not predict the formation of harmonics or multiple simultaneous modes. We would also like to emphasize that the MSHI dispersion relation is based on linear theory and may not capture the full set of physical mechanisms occurring in the nonlinear saturated state of the spoke [14,27].

## VI. Conclusion

In this work we investigated the radial dependence of azimuthally propagating plasma oscillations in an electron beam generated E × B plasma operated with an electron repelling anticathode. Experimental ion probe and fast frame imaging measurements indicate that at least 2 distinct single lobed ($m = 1$) azimuthal modes occurring in the 5-50 kHz frequency band. The findings also indicate the growth of several additional modes in the peripheral plasma region.

The experimental measurements are corroborated with a 2D3V PIC simulation, which is experimentally validated in this work and shows excellent agreement with time averaged radial plasma density, electron temperature, and plasma potential profiles. The azimuthal mode spectrum observed in the PIC simulation corroborates the azimuthal mode spectrum determined experimentally with the ion probe array. The overall agreement with simulation and experiment indicates that the presently considered operating regime is an inherently 2D3V system.

The radial dependence of the anticathode bias on the azimuthal oscillations is analyzed using a 1D electron continuity model. The model indicates that there is enhanced anomalous transport in the peripheral plasma region, which coincides with the growth of the additional azimuthal modes in this region. Furthermore, it is suggested that the suppression of anomalous transport in the central plasma region is caused by a balance of fluxes driven by ionization and electron pressure.

The physical mechanisms leading to the azimuthal mode spectrum are further analyzed by applying the linear theory of MSHI, making consideration of some possible mechanisms by which ion dynamics can affect the instability onset. We find that the lowest frequency $m = 1$ azimuthal mode is closely predicted by the MSHI dispersion relation by neglecting ion dynamics. By considering the radial outflow of ions calculated from solving the ion continuity, we find that the higher frequency mode may be predicted by MSHI with reasonable agreement. However, the novel finding of multiple simultaneously occurring azimuthal modes is not predicted by MSHI with a simple equilibrium ion flow. Future studies should be conducted to investigate the relationship of azimuthal mode formation and resulting ion dynamics, potentially allowing development of e-beam $E \times B$ plasmas with low ion energies in the peripheral plasma region that are more suited for gentle surface processing applications.

**Acknowledgements**

This research was supported by the U.S. Department of Energy, Office of Fusion Energy Science, under Contract No. DEAC02-09CH11466, as a part of the Princeton Collaborative Low Temperature Plasma Research Facility (PCRF). The authors are grateful to Ivan Romadanov, Sunghyun Son, and Igor Kaganovich for fruitful discussions regarding the physics of electron beam generated E×B plasmas and cross field transport and Timothy K. Bennett for technical support on the electron beam chamber.

**Appendix**

1. **Electron energy distribution function and macroscopic plasma parameters**

The EEDFs and macroscopic parameters determine by Langmuir and emissive probes used in this work are shown in Figure 15, Figure 16, and Figure 17.

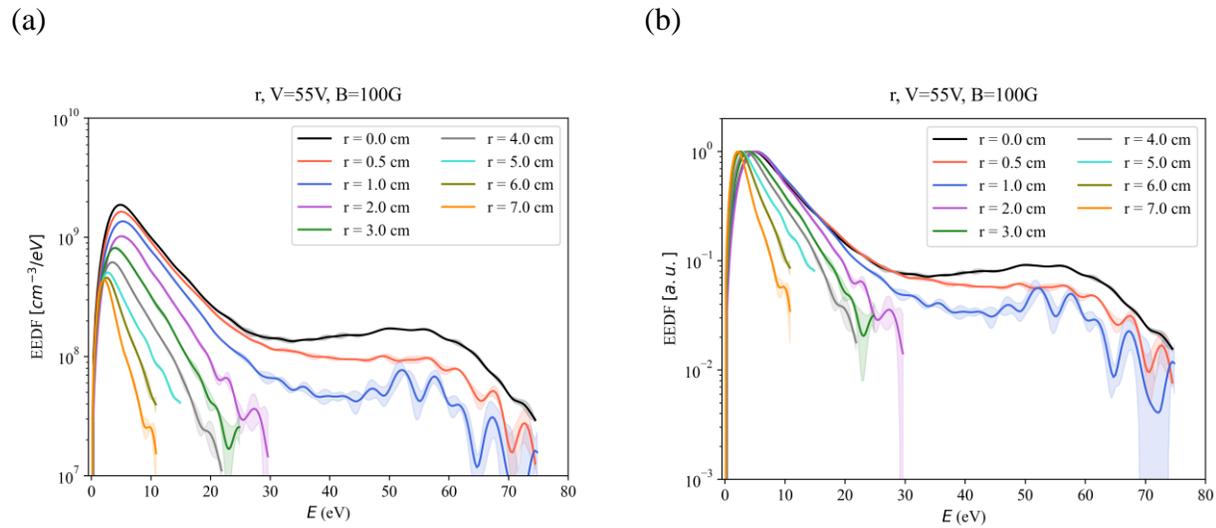

**Figure 15.** EEDF determined by LP at $B = 100$ G, $V_c = 55$ V, $p = 0.1$ mTorr, for (a) in absolute units, (b) normalized to EEDF maximum.

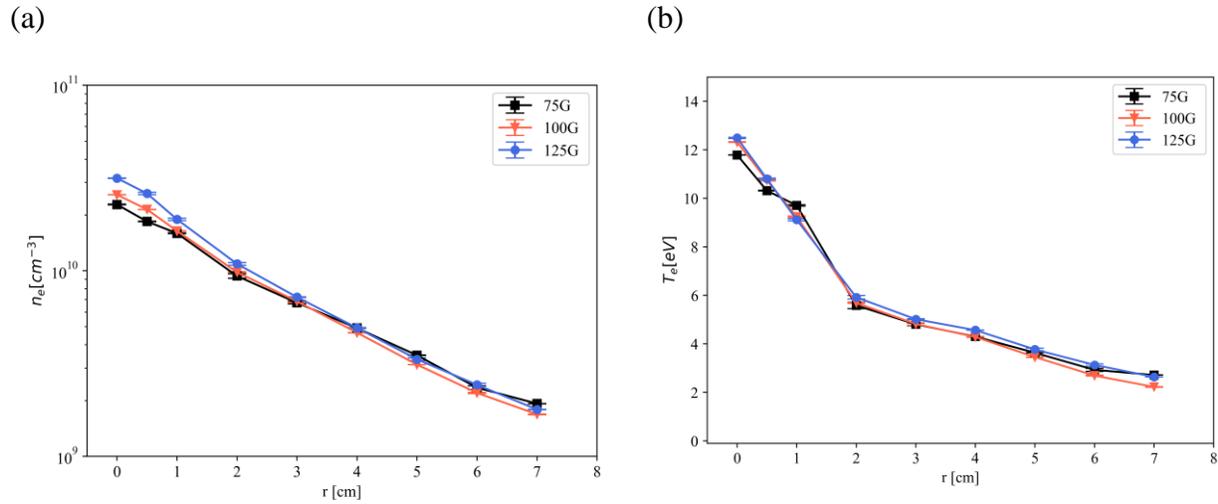

**Figure 16.** Radial profile of (a) plasma density $n_e$ determined by LP and (b) electron temperature $T_e$ for varying magnetic field, $V_c$ = 55 V, $p$ =0.1 mTorr.

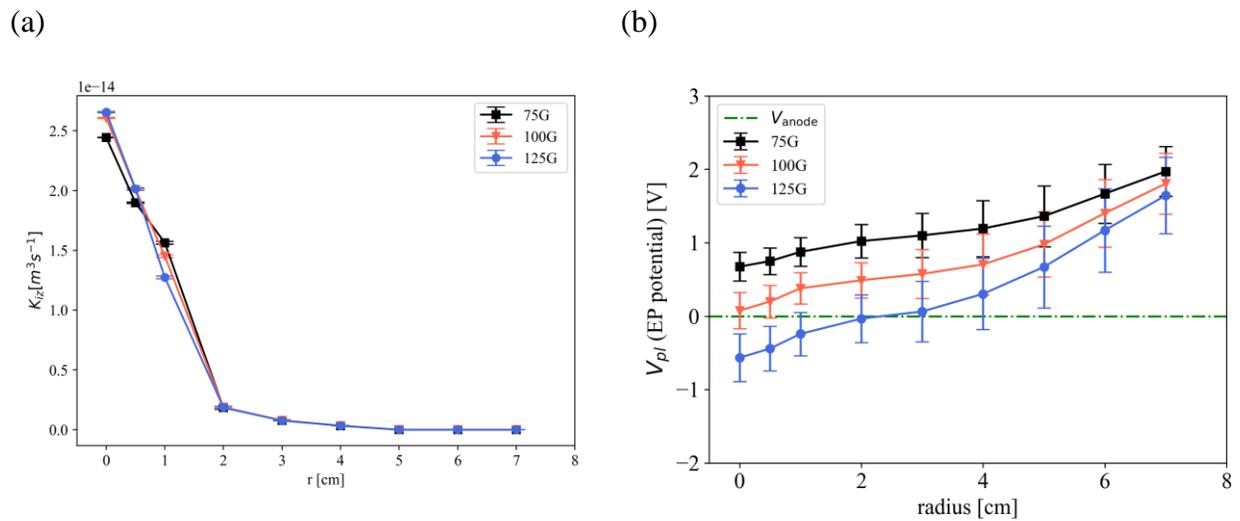

**Figure 17.** (a) Ionization rate constant $K_{iz}$ radial profile determined by LP and (b) plasma potential $V_{pl}$ radial profile determined by EP for varying magnetic field, $V_c$ = 55 V, $p$ =0.1 mTorr.

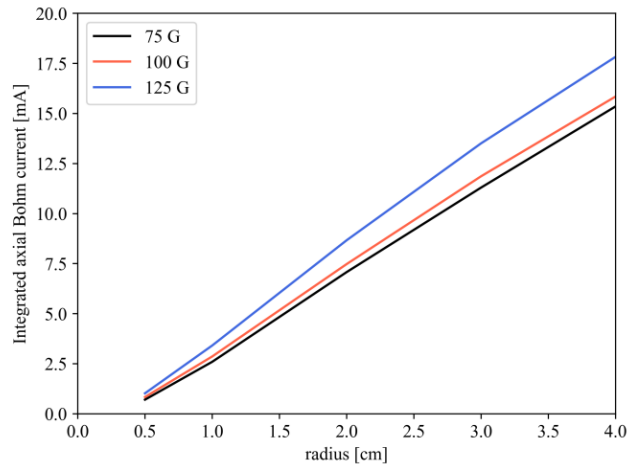

**Figure 18.** Surface integrated Bohm ion current as determined by Eq. (14), for varying magnetic field and parameters $V_c = -55$ V, $p = 0.1$ mTorr.

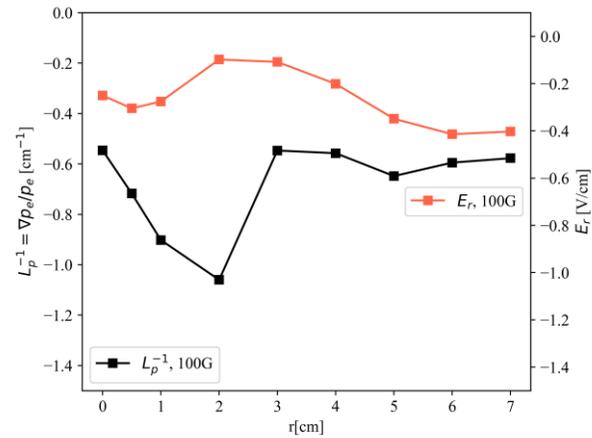

**Figure 19.** Radial profiles of electron pressure gradient inverse length scale $L_p^{-1}$ and radial electric field for. $B = 100$ G, $V_c = -55$ V, $p = 0.1$ mTorr.

2. **Fast frame imaging corroboration of azimuthal plasma oscillations**

An example fast frame image sequence of the optical plasma emission in the $r-z$ plane of the plasma is shown in Figure 20.

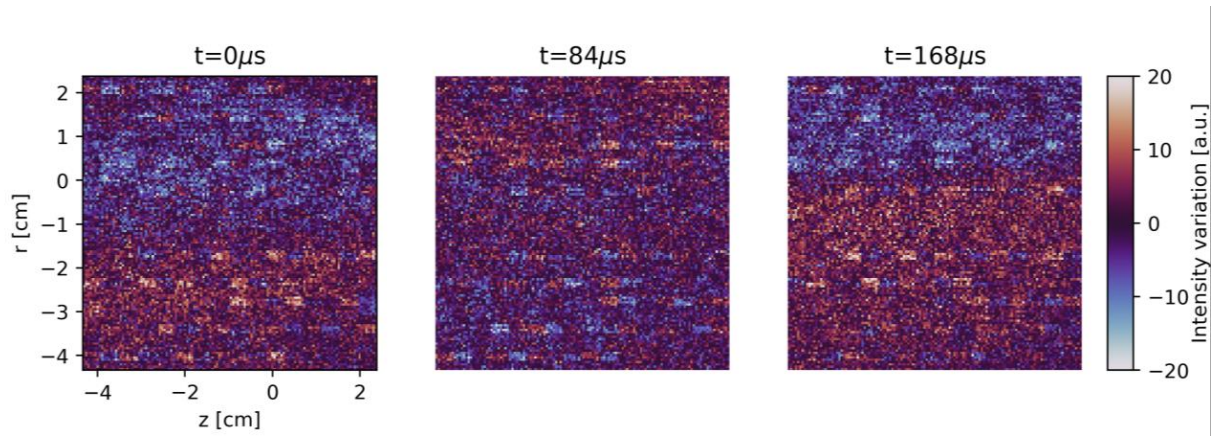

**Figure 20.** Mean subtracted fast frame image sequence of one cycle of the 6 kHz mode. Images are in the $r-z$ plane, with radial coordinate on the vertical axis and axial coordinate on the horizontal axis. The red region indicates an optically brighter region of the plasma, while the blue region indicates optically dimmer, as compared to the time averaged brightness of the fast frame sequence. Experimental parameters are $B = 100$ G, $V_c = -55$V.

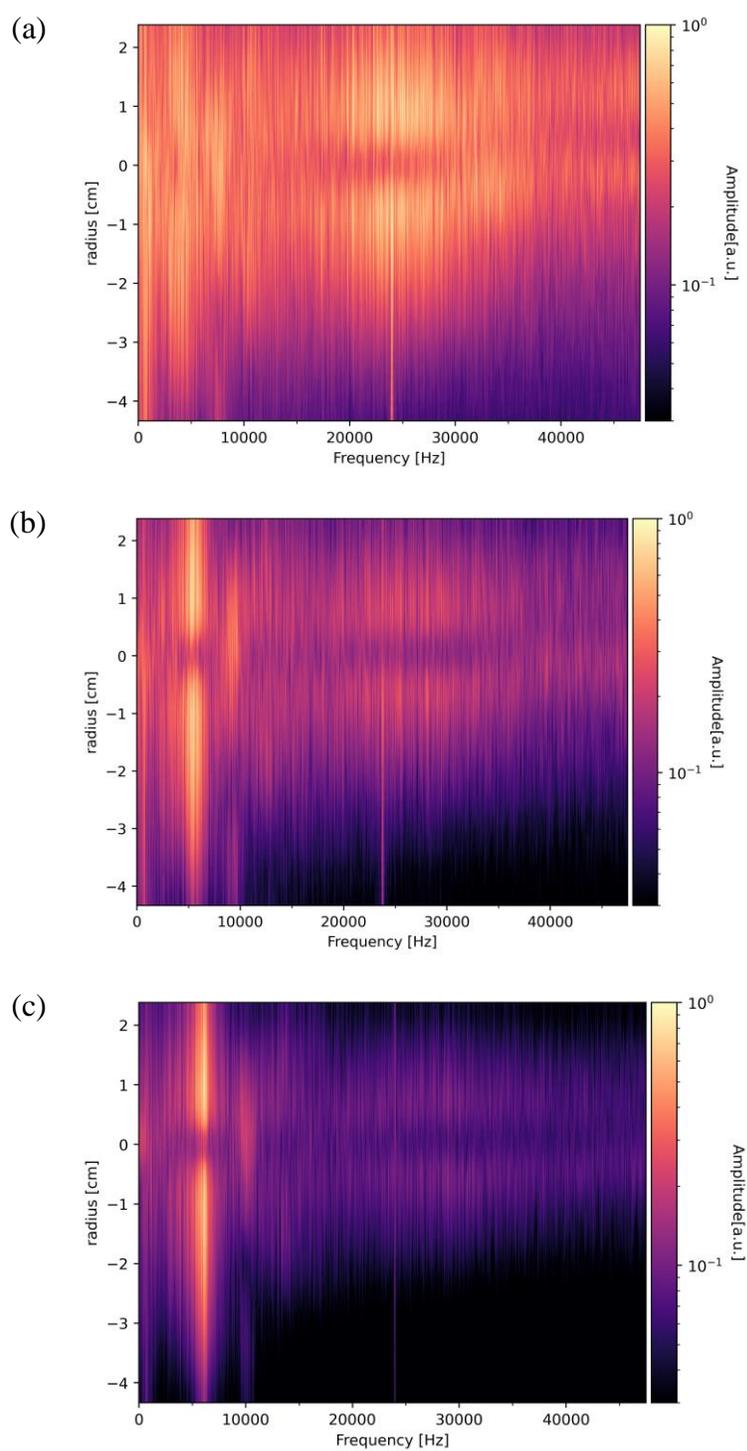

**Figure 21.** Spectrograms of the axially averaged fast frame sequence taken at 95 kfps for magnetic field (a) $B = 75$ G, (b) $B = 100$ G, and (c) $B = 125$ G. Experimental parameters are $p = 0.1$ mTorr, $V_c = -55$V.